\documentclass[twocolumn,aps,superscriptaddress]{revtex4-1}
\usepackage[dvips]{graphicx}
\usepackage{latexsym,amssymb,amsmath}
\usepackage{color}
\usepackage{bm}
\usepackage{enumerate}
\usepackage[bookmarksnumbered,bookmarksopen,colorlinks,citecolor=red,linkcolor=blue]{hyperref}
\usepackage{mathrsfs}
\usepackage{times}
\usepackage{csquotes}
\usepackage{multirow}
\usepackage{feynmp}
\usepackage{tikz}

\graphicspath{{./figs/}}

\begin{document}   

\title{$J/\psi$ polarization in relativistic heavy ion collisions}
\author{Jiaxing Zhao}
\affiliation{SUBATECH, Universit\'e de Nantes, IMT Atlantique, IN2P3/CNRS, 4 rue Alfred Kastler, 44307 Nantes cedex 3, France}
\author{Baoyi Chen}
\affiliation{Department of Physics, Tianjin University, Tianjin 300354, China}
\date{\today}

\begin{abstract}
Understanding the polarization property of $J/\psi$ is critical to constrain its production mechanism. In addition, the polarization of $J/\psi$ can reveal the impact of strong electromagnetic and vorticity fields in relativistic heavy ion collisions. In this study, we analyzed the yield and polarization of $J/\psi$ in relativistic heavy ion collisions at different centrality and transverse momentum regions, using three different reference frames: the Collins-Soper frame, the helicity frame, and the event plane frame. The polarization of initially produced $J/\psi$ is determined by the NRQCD calculation and is similar to that of $pp$ collisions. However, both unpolarization and transverse polarization are considered for the regenerated $J/\psi$.
Our results indicate that the polarization at high $p_T$ is similar to that observed in $pp$ collisions. However, at low $p_T$, where regenerated $J/\psi$ dominates, it is likely that the polarized charm quarks in the rotational QGP medium are responsible for this phenomenon. Our study supplies a baseline for future research on the effects of strong electromagnetic and vorticity fields on $J/\psi$ polarization.
\end{abstract}
\maketitle 

\section{Introduction}

Quarkonium is one of the most ideal probes for quantum chromodynamics (QCD), which provides significant progress in understanding of the QCD in both perturbative and non-perturbative ways. Quarkonium production can be dealt with the factoirzation theory. Heavy quark pairs are produced via perturbative QCD, due to $m_Q\gg \Lambda_{\rm QCD}$, with heavy quark mass $m_Q$ and the QCD cutoff $\Lambda_{\rm QCD}$. The formation of colorless quarkonium from the heavy quark pairs $Q\bar Q$ is a non-perturbative process, which is usually encoded in some parameters, such as the long-distance matrix elements in Non-Relativistic QCD (NRQCD), and it can be fixed by the experimental data.
Quarkonium production and momentum spectrum in $pp$ collisions have been well described by many theoretical models, such as the Color-Evaporation Model (CEM)~\cite{Fritzsch:1977ay,Amundson:1995em}, the Color-Singlet Model (CSM)~\cite{Chang:1979nn,Baier:1981uk} and the Color-Octet Model (COM)~\cite{Bodwin:1992qr}, the latter two are encompassed in the NRQCD~\cite{Bodwin:1994jh,Butenschoen:2012px,Gong:2012ug,Chao:2012iv}. 

However, the yield and spectrum are insufficient to rule out anything so far, and it's hard to get a further understanding of the production mechanism. The quarkonium polarization introduces a new dimension to constrain the production mechanism. For the polarization of $J/\psi$, the CSM gives a transversal polarization at leading order (LO) but a longitudinal polarization at next to leading order (NLO)~\cite{Gong:2008sn}. The COM shows a small longitudinal polarization at small $p_T$ regions~\cite{Beneke:1996tk}, while a transverse polarization at large $p_T$ regions~\cite{Gong:2008ft}. NRQCD includes all important contributions from both color-singlet and color-octet intermediate states and gives a detailed interview of the $J/\psi$ polarization~\cite{Butenschoen:2012px,Gong:2012ug,Chao:2012iv}. Further, NRQCD plus Color Glass Condensate (CGC) makes the prediction down to very low $p_T$ regions~\cite{Ma:2018qvc}. Recently, the polarization of $J/\psi$ is also studied via the improved color evaporation model~\cite{Cheung:2018tvq,Cheung:2021epq}. In experiments, $J/\psi$ polarization has been observed in different colliding systems, energies, and different kinematic ranges~\cite{CDF:2000pfk,LHCb:2013izl,ALICE:2011gej,ALICE:2018crw}. However, due to the large uncertainty, it is hard to get a conclusion so far. 

In relativistic heavy ion collisions, there are two sources for $J/\psi$ production: one is the initial production through hard processes which is largely suppressed in the medium~\cite{Gerschel:1988wn,NA50:1996lag}, and the other one is the regeneration in the quark-gluon plasma (QGP) via the recombination from uncorrelated charm and anticharm~\cite{Thews:2000rj,Grandchamp:2002wp,Yan:2006ve}. The competition between the suppression and regeneration can explain well the experimentally measured $J/\psi$ nuclear modification factor $R_{AA}$ and collective flows~\cite{Du:2015wha,Zhao:2010nk,Zhou:2014kka,Zhao:2021voa,Chen:2018kfo,Liu:2009wza,He:2021zej,Villar:2022sbv} at RHIC and LHC energies. Except for the yield, 
the polarization of $J/\psi$ concerning the event plane in $PbPb$ collisions at $\sqrt{s_{\rm NN}}$ = 5.02 TeV at the LHC is observed in the first time~\cite{ALICE:2022dyy}. Initial produced $J/\psi$ may carry different polarization information compared to the regenerated one, as discussed before~\cite{Ioffe:2003rd}. So, this will give new constraints on the regeneration component and 
also reveal the influence of strong electromagnetic and vorticity fields on the polarization of $J/\psi$.

In this paper, we want to study the yield and polarization of $J/\psi$ in relativistic heavy ion collisions. The paper is organized as follows. The $J/\psi$ polarization in $pp$ collisions is shown in Section~\ref{sec2}. Next the production and transverse spectrum of $J/\psi$ in $PbPb$ collisions at $\sqrt{s_{\rm NN}}$ = 5.02 TeV is investigated in Section~\ref{sec3}. In Section~\ref{sec4}, we will show the polarization of $J/\psi$ in $PbPb$ collisions. A conclusion will be given in Section~\ref{sec5}.

\section{polarization in proton-proton collisions}
\label{sec2}

The geometrical shape of the angular distribution of the two decay products reflects the polarization of the quarkonium state in dilepton decays. 
The distribution of the $J/\psi$ decay products can be expressed in general as
\begin{eqnarray}
W(\theta, \phi)&\propto&{1\over 3+\lambda_\theta} (1+\lambda_\theta \cos^2\theta +\lambda_\phi \sin^2\theta \cos2\phi)\nonumber\\
&+&\lambda_{\theta\phi} \sin2\theta \cos \phi),
\end{eqnarray}
where $\theta$ and $\phi$ are the polar and azimuthal angles in a given reference frame. The polarization parameters are related to the spin states of decay products, so the spin state of $J/\psi$. The polarization is completely longitudinal if the set of polarization parameters $(\lambda_\theta,\lambda_\phi,\lambda_{\theta\phi})$ takes the values $(-1,0,0)$ and it is completely transverse if it takes the values $(1,0,0)$. In the unpolarization scenario, the parameters are $(0,0,0)$.

\begin{figure*}[!htb]
\includegraphics[width=0.8\textwidth]{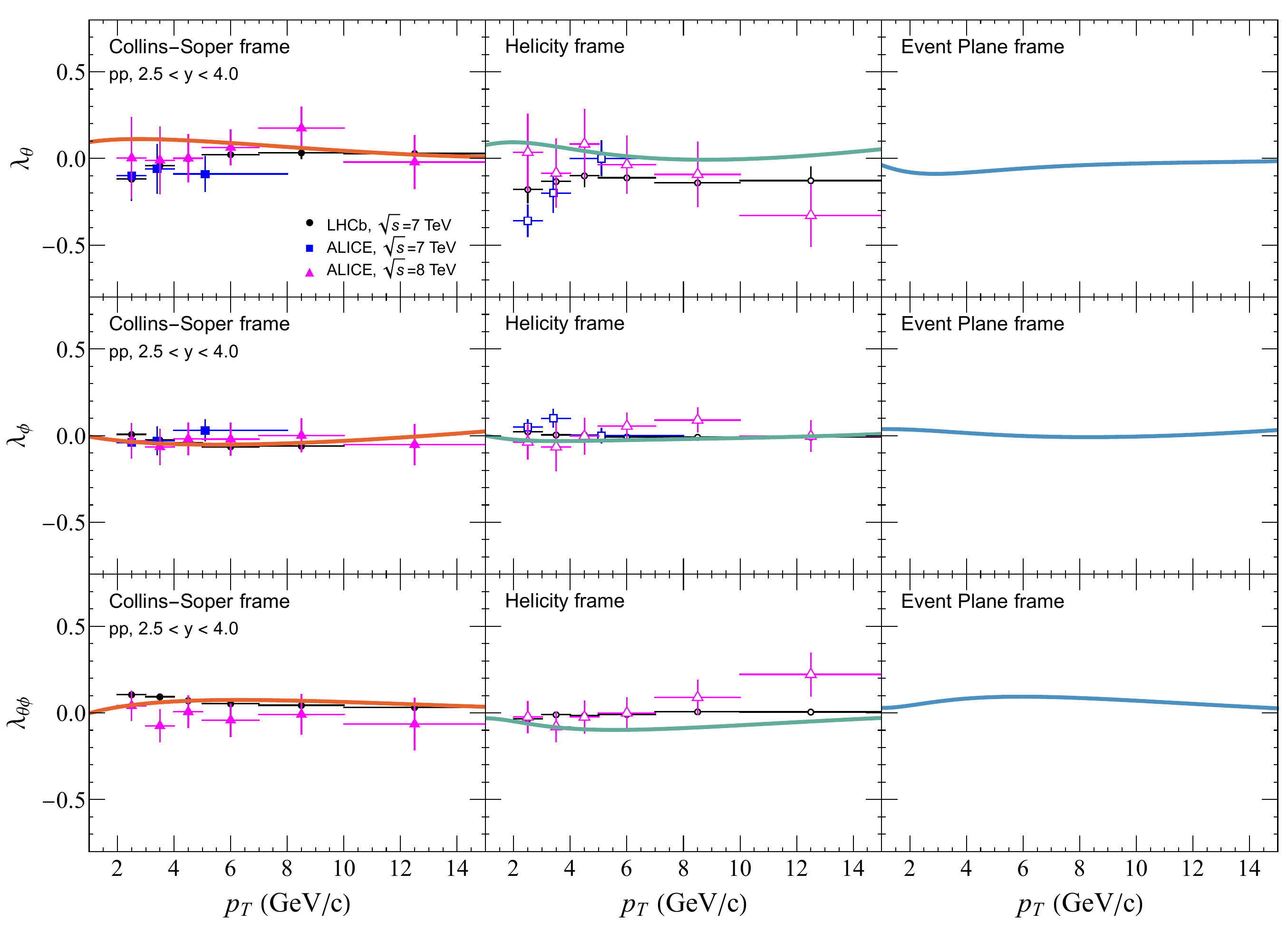}
\caption{$p_T$ dependent $\lambda_\theta$, $\lambda_\phi$, and $\lambda_{\theta\phi}$ in forward rapidity $2.5<y<4.0$ of $pp$ collisions in the Collins-Soper frame (left), the helicity frame (middle), and the event plane frame (right). The experiment data is from LHCb $\sqrt{s}=7\rm TeV$~\cite{LHCb:2013izl}, ALICE $\sqrt{s}=7\rm TeV$~\cite{ALICE:2011gej}, and ALICE $\sqrt{s}=8\rm TeV$~\cite{ALICE:2018crw}. Solid lines in the left and middle panels are from the NRQCD calculation and taken from Ref.~\cite{Ma:2018qvc} directly, while the lines in the right panel are from the transformation as shown in Eq.~\eqref{eq.trans}.}
\label{fig.lambdathe}
\end{figure*}

The polarization parameters are frame-dependent, which is defined with respect to a particular polarization axis, normally taken as $z-$axis. There are three main reference frames. There are the Collins-Soper (CS) frame, the helicity frame (HX), and the newly used event plane frame (EP). In the CS frame, the $z-$axis is defined as the bisector of the angles between the direction of one beam and the opposite of another beam in the rest frame of the $J/\psi$. In the HX frame, $z-$axis is defined as the direction of the $J/\psi$ in the centre of mass frame of two colliding nucleus, which is same as the Laboratory frame for the colliding system. In the EP frame, $z-$axis is defined as the direction orthogonal to the event plane in the centre of mass frame of two colliding nucleus. This is also the direction of the strongest net-magnetic field and global vorticity field in heavy ion collisions. 

Although the polarization parameters $(\lambda_\theta,\lambda_\phi,\lambda_{\theta\phi})$ depend on the frame, the values in different frame are connected via the geometrical transformation. The polarization parameters between two reference frames are given by~\cite{Faccioli:2010kd},
\begin{eqnarray}
\lambda'_\theta&=&{\lambda_\theta-3\Lambda \over 1+\Lambda},\nonumber\\
\lambda'_\phi&=&{\lambda_\phi+\Lambda \over 1+\Lambda},\nonumber\\
\lambda'_{\theta\phi}&=&{\lambda_{\theta\phi}\cos2\delta-(\lambda_\theta-\lambda_\phi)/2\sin 2\delta \over 1+\Lambda},
\label{eq.trans}
\end{eqnarray}
where $\Lambda=(\lambda_\theta-\lambda_\phi)/2\sin^2\delta-\lambda_{\theta\phi}/2\sin 2\delta$ with the angle $\delta$ between two polarization axis. Usually, this transformation is with a strong kinematic dependence, which is encoded in the angles $\delta$. The angle between the CS and HX frame can be given,
\begin{eqnarray}
\delta_{\rm CS\to HX}=\arccos \left(m_{J/\psi}\sinh y\over \sqrt{p_T^2+E_T^2\sinh^2 y} \right),
\end{eqnarray}
where $y$ is the rapidity and $p_T$ is transverse momentum. The transverse energy $E_T=\sqrt{m_{J/\psi}^2+p_T^2}$.
Based on the definition, the angle between the HX and EP frame can be given by
\begin{eqnarray}
\delta_{\rm HX\to EP}=\arccos \left(p_T\sin\phi \over \sqrt{p_T^2+E_T^2\sinh^2 y} \right).
\end{eqnarray}
So it depends not only on the $J/\psi$ momentum $p_T$ but also the angle $\phi$ in the transverse plane. The angle $\phi$ dependent distribution can be expressed as the anisotropic flow. Such as the elliptic flow, its definition is $v_2\equiv \langle \cos 2\phi \rangle$. So, $\langle \phi\rangle=1/2\arccos v_2$. The angles $\delta$ are related to the value of the $v_2$. If the $v_2=0$ for any $p_T$ regions, we obtain $\langle \phi \rangle=\pi/4$. So, the $\delta_{\rm HX\to EP}\approx \pi/2$ in the forward rapidity. 

The experimental data of $J/\psi$ $\lambda_\theta$, $\lambda_\phi$, and $\lambda_{\theta\phi}$ are shown in Fig.~\ref{fig.lambdathe}. The solid lines in the CS and HX frame are from the NRQCD calculation~\cite{Ma:2018qvc}, which are related to each other via the transformation shown in Eq.~\eqref{eq.trans}. The values in the EP frame, which is shown with blue lines in the right panel, are obtained from the HX frame quantities via the above transformation and the angle $\delta_{\rm HX\to EP}$.

\section{$J/\psi$ transport in heavy ion collisions}
\label{sec3}

We now focus on charmonium transport and production in heavy ion collisions, which can be described by a classical transport equation~\cite{Du:2015wha,Zhao:2010nk,Zhou:2014kka,Zhao:2021voa,Chen:2018kfo,Liu:2009wza,He:2021zej}. The charmonium phase space distribution $f_\psi({\bm p},{\bm x},\tau|{\bm b})$, where $\psi=J/\psi, \chi_c, \psi'$, is governed by the Boltzmann equation~\cite{Zhao:2020jqu},
\begin{eqnarray}
\label{transport}
&&\left[ \cosh(y-\eta)\partial_\tau + {\sinh(y-\eta)\over \tau}\partial_\eta+{\bm v}_T\cdot \nabla_T \right] f_\psi\nonumber\\
= && -\alpha f_\psi +\beta,
\end{eqnarray}
where $y=1/2\ln[(E+p_z)/( E-p_z)]$ is the charmonium momentum rapidity, and ${\bm v}_T={\bm p}_T/E_T$ is the charmonium transverse velocity with transverse energy $E_T$. The second and third terms on the left-hand side arise from the free streaming of $\psi$, which leads to the leakage effect in the longitudinal and transverse directions. The charmonium suppression and regeneration processes in the QGP medium are reflected in the loss term $\alpha$ and gain term $\beta$ in the right-hand side, respectively.

In the hot QGP medium, the interaction potential between charm and anticharm suffers Debye screening~\cite{Matsui:1986dk}. This may leads to the suppression of the charmonium states. The dissociation temperatures $T_d$ of charmonium states can be given by solving the Schr\"odinger equation with the temperature-dependent heavy quark potential. In this study, we take $T_d\approx 2.3T_c$ for $J/\psi$, which can be obtained by using the potential from lattice QCD~\cite{Lafferty:2019jpr} as well as the internal energy~\cite{Satz:2005hx}. Besides the Debye screening, charmonium may also suffer the dynamical dissociation in the QGP via scattering with thermal partons, such as the gluon dissociation~\cite{Peskin:1979va,Bhanot:1979vb} $g\psi \to c \bar c$. For the ground state $J/\psi$, the gluon dissociation cross-section $\sigma_{g\psi}^{c\bar c}$ in vacuum is derived using the operator-production-expansion (OPE) method~\cite{Peskin:1979va,Bhanot:1979vb}. For the excited states $\chi_c$ and $\psi'$, the cross sections can be obtained through their geometric relation to the ground state~\cite{Chen:2018kfo}. In this study, we take only the gluon dissociation as the loss term and its inverse process $c\bar c\to g\psi$ as the gain term, $\alpha$, and $\beta$ can be explicitly expressed as~\cite{Zhao:2020jqu}
\begin{eqnarray}
\label{alphabeta}
\alpha({\bm p},{\bm x},\tau|{\bf b}) &=& {1\over 2E_T}\int{d^3{\bm p}_g \over (2\pi)^3 2E_g}W_{g\psi}^{c\bar c}(s)f_g({\bm p}_g,{\bm x},\tau)\nonumber\\
&&\times\Theta(T({\bm x},\tau|{\bm b})-T_c),\nonumber\\
\beta({\bm p},{\bm x},\tau|{\bm b}) &=& {1\over 2E_T}\int {d^3{\bm p}_g \over (2\pi)^3 2E_g}{d^3{\bm p}_c \over(2\pi)^3 2E_c}{d^3{\bm p}_{\bar c} \over(2\pi)^3 2E_{\bar c}}\nonumber\\
&&\times W_{c\bar c}^{g\psi}(s)f_c({\bm p}_c,{\bm x},\tau|{\bm b})f_{\bar c}({\bm p}_{\bar c},{\bm x},\tau|{\bm b})\nonumber\\
&&\times(2\pi)^4\delta^{(4)}(p+p_g-p_c-p_{\bar c})\nonumber\\
&&\times\Theta(T({\bm x},\tau|{\bm b})-T_c),
\end{eqnarray}
where $E_g, E_c$, and $E_{\bar c}$ are the gluon, charm quark, and anti-charm quark energies, ${\bm p}_g, {\bm p}_c$ and ${\bm p}_{\bar c}$ are their momenta, and $s$ is the invariant mass of $g\psi$ system. $\bm b$ is impact parameter of two nucleus. $W_{g\psi}^{c\bar c}$($W_{c\bar c}^{g\psi}$) is the dissociation(regeneration) probability. The regeneration probability and dissociation probability are connected via the detailed balance~\cite{Yan:2006ve,Liu:2009wza}. The step function $\Theta$ is used to guarantee the calculation in the QGP phase above the critical temperature $T_c$, and the local temperature of the medium $T({\bm x},\tau)$ is given by the hydrodynamic simulation. In this paper, we employ a (2+1)-dimensional hydrodynamic model, the MUSIC package~\cite{Schenke:2010nt,McDonald:2016vlt}, to characterize the space and time dependence of the temperature and velocity of the hot medium. 

The gluon distribution $f_g$ is chosen as the Bose-Einstein distribution in the loss term. The experimentally measured large charmed meson flow indicates that charm quarks with small transverse momentum might be thermalized in the medium~\cite{STAR:2017kkh}. So, as a first approximation, we take a thermal distribution for charm (anti-charm) quarks in the momentum space $f_c({\bm p}_c,{\bm x})=\rho_c({\bm x}) N_c/(e^{p_c^\mu u_\mu/T}+1)$, where $N_c$ is the normalization factor, and the density in coordinate space $\rho_c$ is controlled by the charm conservation equation $\partial_\mu(\rho_cu^\mu)=0$. The initial density at time $\tau_0$ is governed by the nuclear geometry of the colliding system,  
\begin{equation}
\label{charm}
\rho_c({\bm x},\tau_0|{\bm b})={T_A({\bm x}_T+{\bf b}/2)T_B({\bm x}_T-{\bm b}/2)\cosh \eta \over \tau_0}{d\sigma_{pp}^{c\bar c}\over d\eta},
\end{equation}
where $T_A$ and $T_B$ are the thickness functions~\cite{Miller:2007ri}. ${\bm x}_T$ is transverse radius. $d\sigma^{c\bar{c}}_{pp}/d\eta$ is the rapidity distribution of charm quark production cross section in $pp$ collisions. We take 0.43 mb $< d\sigma^{c\bar{c}}_{pp}/d\eta<$ 0.61 mb in the forward rapidity of $PbPb$ collisions at $\sqrt{s_{\rm NN}}=5.02$ TeV with considering the shadowing effect~\cite{LHCb:2016ikn}. 

\begin{figure}[!htb]
\includegraphics[width=0.35\textwidth]{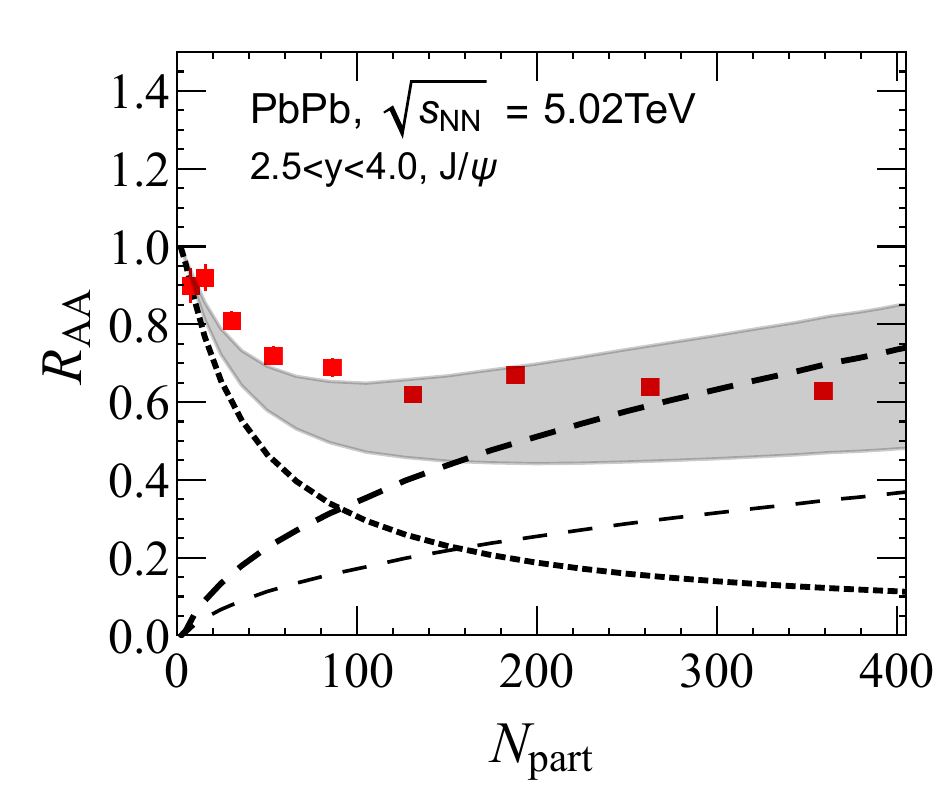}\\
\includegraphics[width=0.35\textwidth]{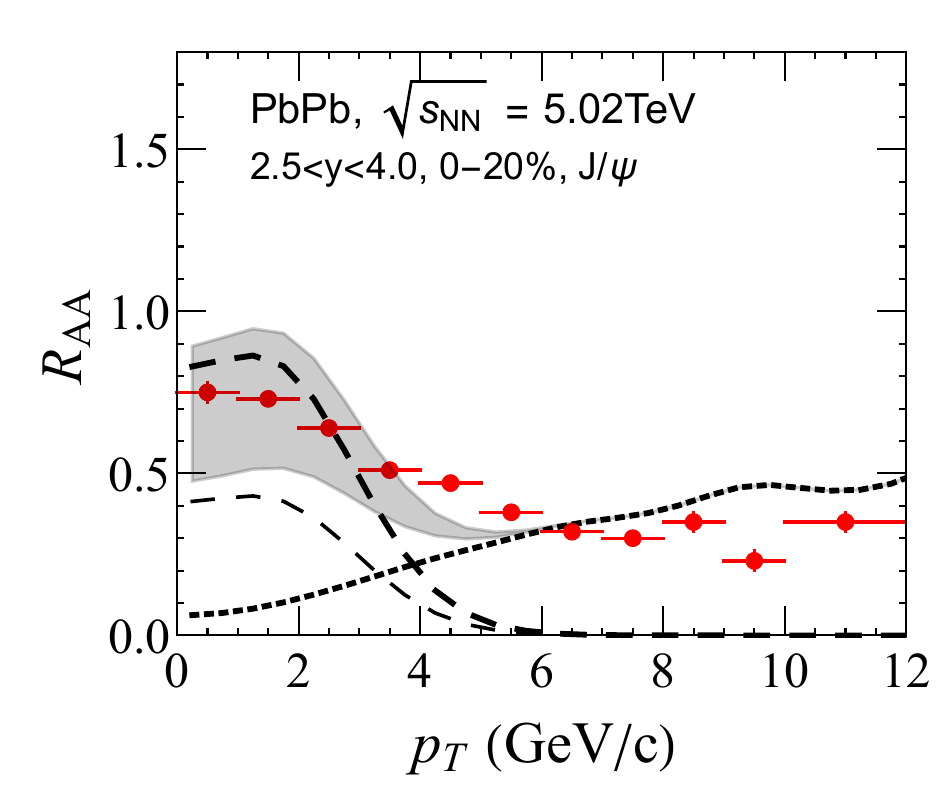}
\caption{$J/\psi$ $R_{AA}$ in forward rapidity $2.5<y<4$ of $PbPb$ collisions with $\sqrt{s_{\rm NN}}=5.02 \rm TeV$ as a function of number of participant $N_{\rm part}$  (upper) and transverse momentum $p_T$ (lower) in 0-20\% centrality regions. The dotted and dashed lines represent initial and regeneration contributions, respectively. The thick and thin dashed lines are the regeneration fraction with the charm cross section $d\sigma^{c\bar{c}}_{pp}/d\eta=0.61\rm mb$ and 0.43 mb. The bands are the full result with different charm cross sections. The experimental data are from ALICE~\cite{ALICE:2023gco,ALICE:2019lga}.}
\label{fig.raa}
\end{figure}
To solve the transport equation, the initial charmonium distribution is needed, which can be written as
\begin{eqnarray}
\label{eq.initial}
&& f_\psi ({\bm p}, {\bm x}, \tau_0|{\bm b})\nonumber\\
= && {(2\pi)^3 \over E_T \tau_0}\int dz_A dz_B \rho_A( {\bm x}_T+{{\bm b}/2},z_A)\rho_B( {\bm x}_T-{{\bm b}/2},z_B)\nonumber\\
&&\times\mathcal{R}_g(x_1,\mu_F, {\bm x}_T+{\bm b}/2)\mathcal{R}_g(x_2,\mu_F, {\bm x}_T-{\bm b}/2)\nonumber\\
&&\times f_{pp}^{\psi}({\bm x}_T,{\bm p},z_A,z_B|{\bm b}).
\end{eqnarray}
It is just a superposition of the charmonium distribution in $pp$ collisions with the cold nuclear effects, such as nuclear shadowing~\cite{Mueller:1985wy} and Cronin effect~\cite{Cronin:1974zm,Hufner:1988wz}. For heavy ion collisions at LHC energy, the collision time is very short, the nuclear absorption can be safely neglected~\cite{Gerschel:1988wn}.
$\rho_A$ and $\rho_B$ are the nucleon distributions in the two colliding nuclei. The shadowing effect is embedded in the factor $\mathcal R_g$, which is taken from the EPS09 package~\cite{Helenius:2012wd}. Before two gluons fuse into a charmonium, they may acquire additional transverse momentum via multi-scattering with the surrounding nucleons, and this extra momentum would be inherited by the produced charmonium, this is the so-called Cronin effect. 
In this study, we replace the averaged transverse momentum square by $\langle p_T^2\rangle_{pp}+a_{gN}l$, 
where the Cronin parameter $a_{gN}$ is the averaged charmonium transverse momentum square obtained from the gluon scattering with a unit of length of nucleons, and $l$ is the mean trajectory length of the two gluons in the projectile and target nuclei before $c\bar c$ formation. We take $a_{gN} = 0.15\text{GeV}^2/\text{fm}$ for $PbPb$ collisions at $\sqrt{s_{\rm NN}}=5.02 \rm TeV$. $\langle p_T^2\rangle_{pp}$ is the mean transverse momentum square, which can be obtained together with the charmonium momentum distribution $f_{pp}^{\psi}$ by fitting the $pp$ collisions, see detail~\cite{Zhou:2014kka,Chen:2018kfo,Zhao:2021voa}. 
 
The transport equation with the initial distribution \eqref{eq.initial} can be solved analytically, as shown in detail~\cite{Zhou:2014kka,Chen:2018kfo,Zhao:2021voa}. To obtain the prompt $J/\psi$, which is normally measured in the experiment, the feed-down from the excited charmonium states contribution. 
The excited states decay into the ground state with the branching ratio $\mathcal{B}_{\chi_c \to J/\psi}=22\%$ and $\mathcal{B}_{\psi(2S)\to J/\psi}=61\%$ respectively after they moves out of the hot medium~\cite{Workman:2022ynf}. 
The nuclear modification factor $R_{AA}$ as a function of the number of participants $N_{\rm part}$ and transverse momentum $p_T$ are studied and shown in Fig.~\ref{fig.raa}. We can see the $J/\psi$ suffers a large suppression in central collisions, especially the low $p_T$ $J/\psi$. While the regeneration play an important role in low $p_T$ and central collisions, which compensates the suppression and together give a good description of the experimental data. 

\section{polarization in heavy ion collisions}
\label{sec4}
The polarization of $J/\psi$ produced in relativistic heavy ion collisions can be expressed as
\begin{eqnarray}
\lambda_\theta(p_T)={N_{\rm ini}(p_T)\lambda_\theta^{\rm ini}(p_T) + N_{\rm reg}(p_T)\lambda_\theta^{\rm reg}(p_T) \over N_{\rm ini}(p_T)+ N_{\rm reg}(p_T)},
\end{eqnarray}
where $\lambda_\theta^{\rm ini}$ and $\lambda_\theta^{\rm reg}$ are for initial and regenerated $J/\psi$, respectively. The same relationship holds for $\lambda_\phi$ and $\lambda_{\theta \phi}$. 
\begin{figure}[!htb]
$$\includegraphics[width=0.24\textwidth]{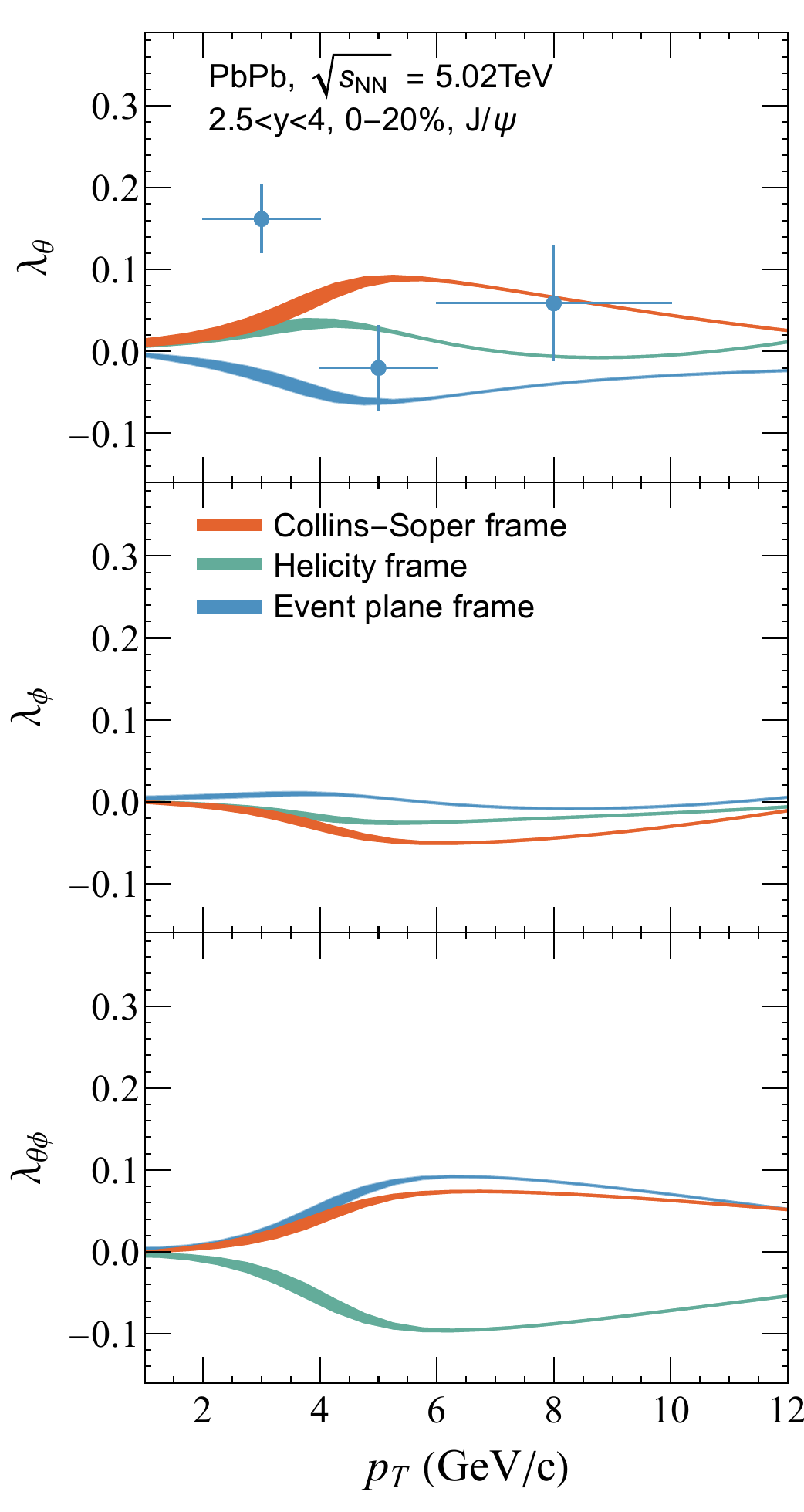}\includegraphics[width=0.24\textwidth]{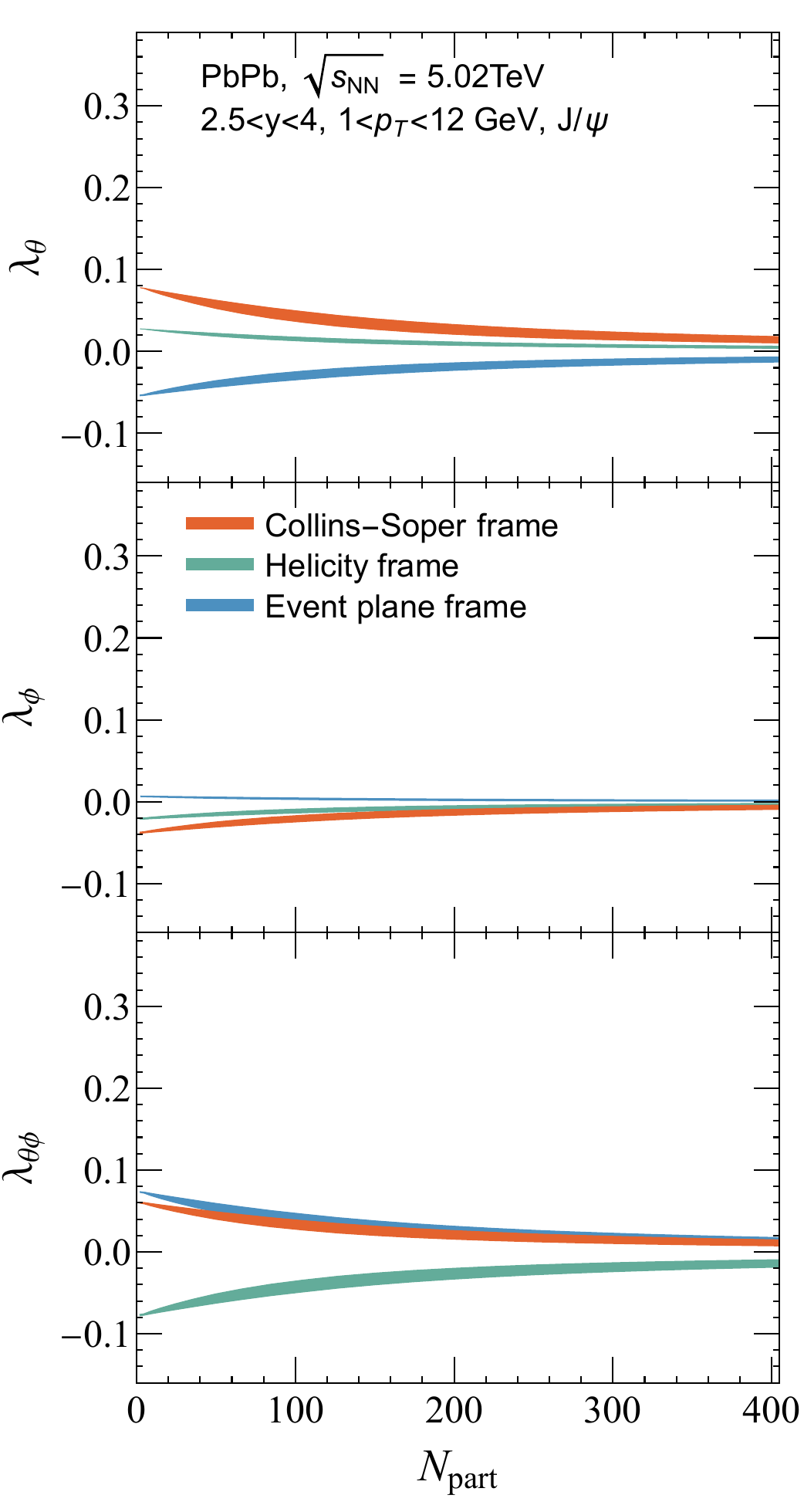}$$
\caption{$p_T$ (left) and $N_{\rm part}$ (right) dependent $J/\psi$ $\lambda_\theta$, $\lambda_\phi$, and $\lambda_{\theta\phi}$ in forward rapidity $2.5<y<4$ of $PbPb$ collisions with $\sqrt{s_{\rm NN}}=5.02 \rm TeV$. The results are shown in the Collins-Soper frame (orange), the helicity frame (green), and the event plane frame (blue). The band is caused by the uncertainty of the nuclear shadowing effect. The experimental data is from ALICE~\cite{ALICE:2022dyy} and obtained concerning the event plane.}
\label{fig.lambda_woreg}
\end{figure}

First, we assume the regenerated $J/\psi$ without polarization information, $\lambda_\theta^{\rm reg}=0$. The polarization of final $J/\psi$ comes completely from the initially produced ones, which is the same as the $pp$. Due to the small interaction cross section and the heavy flavor symmetry, the spin flips of the heavy quark in the hot medium are largely suppressed. The hot medium may melt the initial produced $J/\psi$, but doesn't wash out the polarization of the survived $J/\psi$. With the initial production fraction, we calculate the $p_T$ and number of collisions $N_{\rm part}$ dependent polarization parameters $\lambda_\theta$, $\lambda_\phi$, and $\lambda_{\theta\phi}$ in $PbPb$ collisions within three different frames as shown in Fig.~\ref{fig.lambda_woreg}. 
Because high $p_T$ $J/\psi$, e.g. $p_T>5\rm GeV$, almost comes from the initial production, as shown in Fig~\ref{fig.raa}, their polarization should be the same as the initial production. It's indicated by the experimental data (blue points) and shown in Fig.~\ref{fig.lambda_woreg} that the $\lambda_\theta$ at $p_T>5\rm GeV$ is close to the theoretical calculation (blue line). For the centrality dependence, we can see the polarization disappear at central collisions in this case due to the regeneration dominates. In the very peripheral collisions, the polarization is consistent with the $pp$ collisions.

\begin{figure}[!htb]
$$\includegraphics[width=0.24\textwidth]{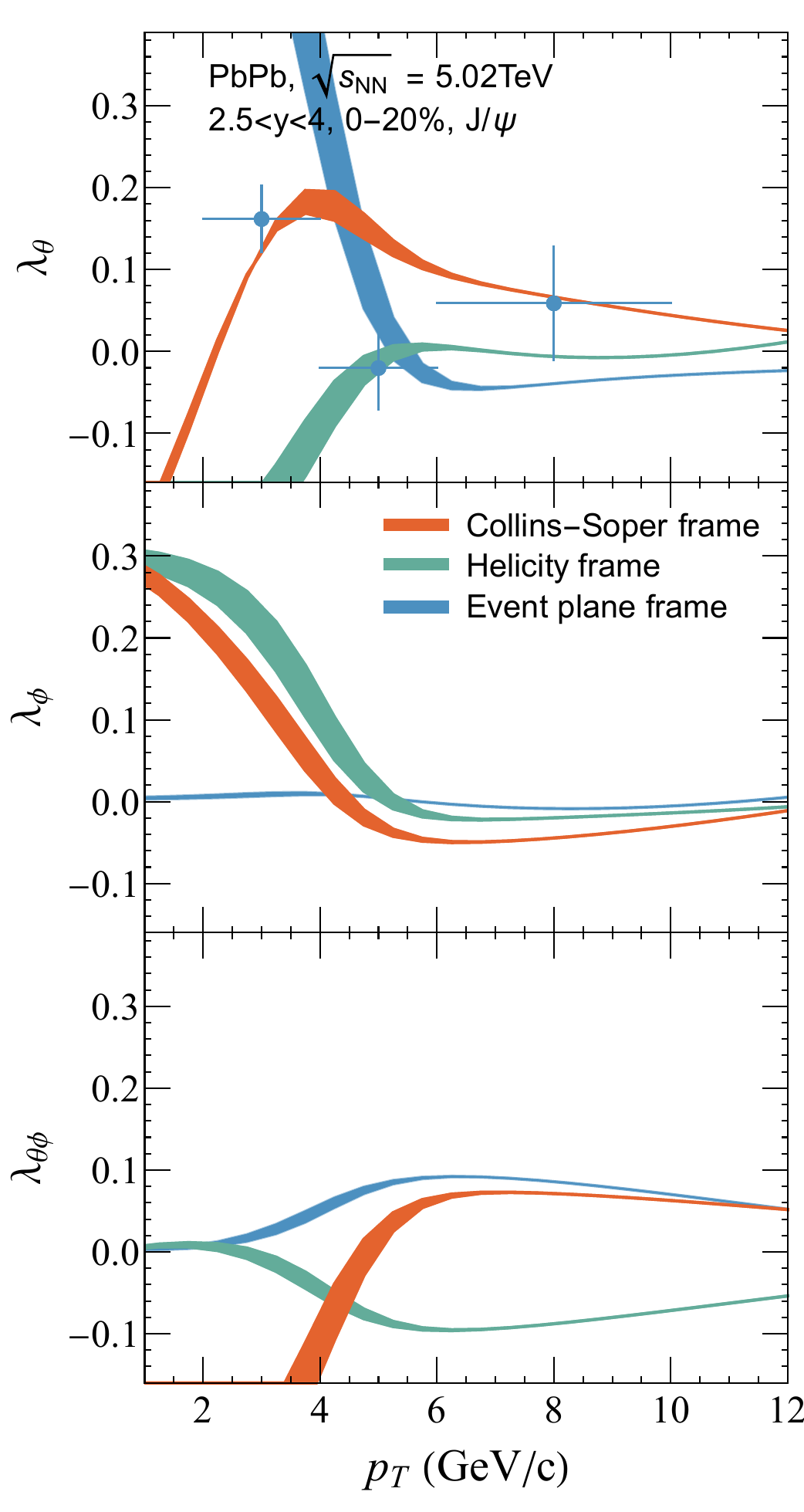}\includegraphics[width=0.24\textwidth]{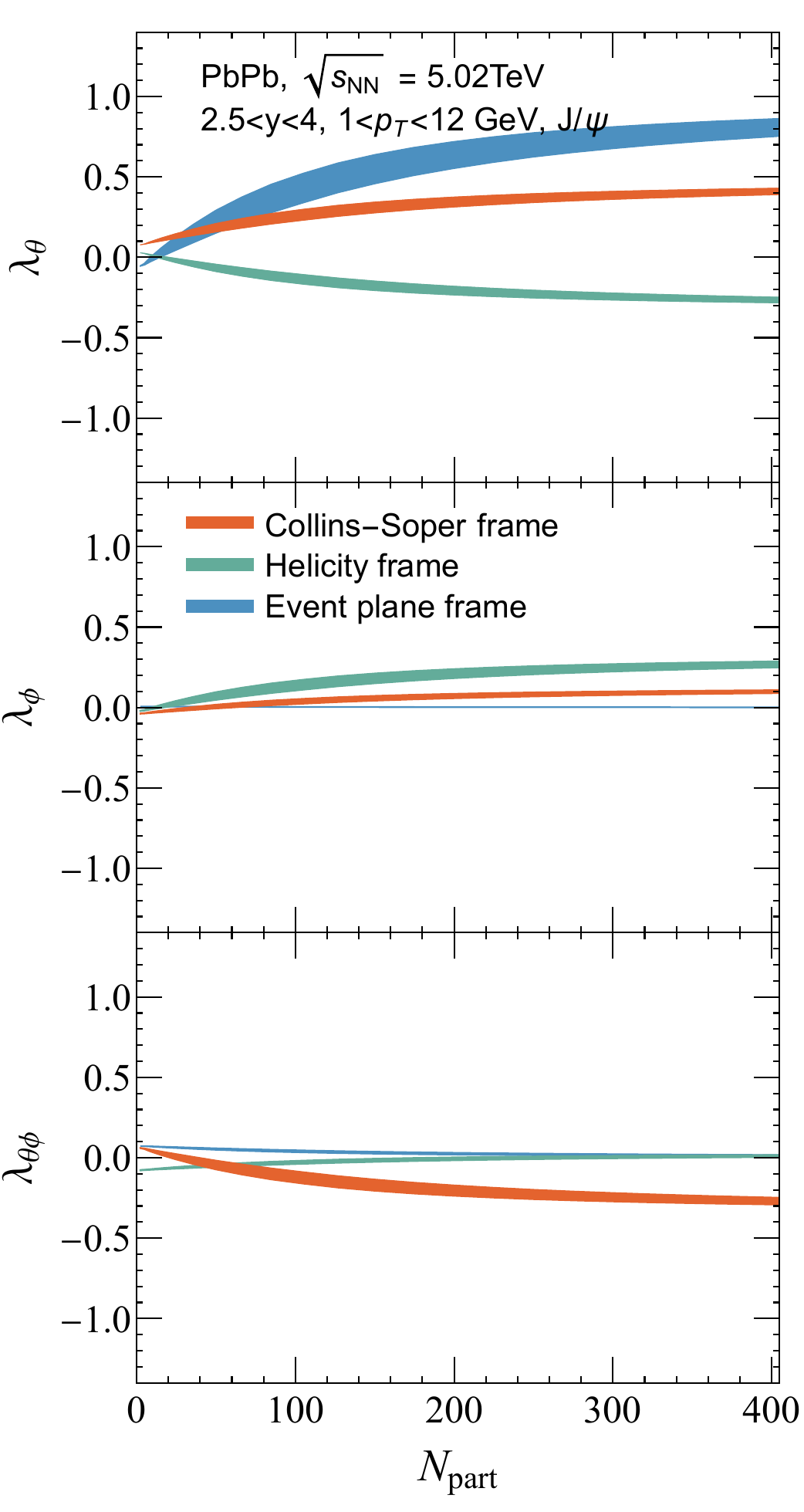}$$
\caption{Same as the Fig.~\ref{fig.lambda_woreg}, but considering the polarization of the regenerated $J/\psi$.}
\label{fig.lambda_reg}
\end{figure}

For the low $p_T$, we can see the larger difference between the theoretical result and experimental data. This indicates the contribution of the regenerated $J/\psi$ cannot be ignored. The polarization of the regenerated $J/\psi$ is from the polarized and uncorrelated charm and anticharm quarks in the QGP. The charm quark can be polarized via coupling to the electromagnetic and vorticity field as light quark~\cite{Liang:2004xn,Becattini:2013vja,Yang:2017sdk}. The spin of charm and aticharm quarks will tend towards the axis of rotation due to the coupling ${\bm \omega}\cdot {\bm s}$. In a strong magnetic field, the charm move towards the axis of the magnetic field while the anticharm move towards the opposite direction due to the coupling ${\bm \mu} \cdot {\bm B}$ with the magnetic moment ${\bm \mu}$.     
In the limit, the vorticity field will induce a transverse polarization of $J/\psi$ with $\lambda_\theta=1$, which is combined by the polarized charm and anticharm quarks, while the magnetic field will induce a longitudinal polarization with $\lambda_\theta=-1$. Because the strength of the magnetic field decays fast and becomes very small when $J/\psi$ regenerated~\cite{Voronyuk:2011jd,Inghirami:2016iru,Yan:2021zjc,Wang:2021oqq,Sheng:2019kmk}, we only consider the polarization caused by the vorticity field.
We can see if the regenerated $J/\psi$ is transversely polarized, the $\lambda_\theta$ at low $p_T$ has a large and positive value in the event plane frame, as shown in Fig.~\ref{fig.lambda_reg}. So, the large $\lambda_\theta$ at $3\rm GeV$ observed in the experimental data indicates the largely polarized charm quarks in the QGP.
The high $p_T$ behavior is taken over by the initial production and the same as Fig.~\ref{fig.lambda_woreg}. 

\section{Summary}
\label{sec5}

In summary, we investigated the effects of the hot medium on the production and polarization of $J/\psi$ in relativistic heavy ion collisions across various transverse momentum bins. The polarization observables are presented in three different reference frames: the Collins-Soper frame, the helicity frame, and the event plane frame considering two cases. One is that the polarization of $J/\psi$ in $PbPb$ collisions is assumed to arise from the initial process, which is given by the NRQCD framework, while the charm quarks and regenerated $J/\psi$ are unpolarized. The suppression of the initial $J/\psi$ is attributed to the dissociation of the primordial polarized $J/\psi$ due to color screening effects and parton inelastic scatterings. However, the polarization information of the survived $J/\psi$ doesn't be washed out in the QGP. In the second case, the initial produced is the same as before, while the regenerated $J/\psi$ is transversely polarized, which comes from the polarized charm and anticharm quarks in the vorticity field.  
Compared to the experimental data, we find the polarization of high $p_T$ $J/\psi$ comes from the initial production, and a large $\lambda_\theta$ at low $p_T$ indicates the polarization of the charm quarks in a rotational QGP medium.  
The precise measurement at low $p_T$ will give the constraint of the contribution of the vorticity field and regeneration. 
Additionally, the polarization of charmonium excited states may exhibit differences compared to the ground state. Complete treatment of these processes will be performed in the future.

\vspace{1cm}
\noindent {\bf Acknowledgement}: This work is funded by the European Union’s Horizon 2020 research and innovation program under grant agreement No. 824093 (STRONG-2020). Baoyi Chen is funded by the National Natural Science Foundation of China (NSFC) under Grants No. 12175165.

\bibliographystyle{apsrev4-1.bst}
\bibliography{Ref}

\end{document}